\documentclass[sigconf]{acmart}

\AtBeginDocument{%
  \providecommand\BibTeX{{%
    \normalfont B\kern-0.5em{\scshape i\kern-0.25em b}\kern-0.8em\TeX}}}

\usepackage{subfigure}
\usepackage{xspace}
\usepackage{multirow}
\usepackage{amsmath}
\usepackage{amsfonts}
\usepackage{bm}
\usepackage{soul}
\usepackage{algorithm}
\usepackage{algpseudocode}
\usepackage{booktabs}
\usepackage[normalem]{ulem}        
\usepackage[switch]{lineno}

\newcommand{\eat}[1]{}

\newcommand{\eg}{\emph{e.g.},\xspace}
\newcommand{\ie}{\emph{i.e.},\xspace}

\newcommand{\aka}{\emph{a.k.a.}\xspace}
\newcommand{\etc}{\emph{etc.}\xspace}

\newcommand\figref[1]{Figure~\ref{#1}}

\newcommand\tabref[1]{Table~\ref{#1}}
\newcommand\secref[1]{Section~\ref{#1}}

\newtheorem{defi}{Definition}

\newcommand{\spred}{{\sc JiZhi}\xspace}
\newcommand{\legacy}{{\sc Legacy}\xspace}
\AtBeginDocument{%
  \providecommand\BibTeX{{%
    \normalfont B\kern-0.5em{\scshape i\kern-0.25em b}\kern-0.8em\TeX}}}

\copyrightyear{2021}
\acmYear{2021}
\setcopyright{acmcopyright}\acmConference[KDD '21]{Proceedings of the 27th ACM SIGKDD Conference on Knowledge Discovery and Data Mining}{August 14--18, 2021}{Virtual Event, Singapore}
\acmBooktitle{Proceedings of the 27th ACM SIGKDD Conference on Knowledge Discovery and Data Mining (KDD '21), August 14--18, 2021, Virtual Event, Singapore}
\acmPrice{15.00}
\acmDOI{10.1145/3447548.3467146}
\acmISBN{978-1-4503-8332-5/21/08}
\begin{document}
\fancyhead{}

\title{\spred: A Fast and Cost-Effective Model-As-A-Service System for Web-Scale Online Inference at Baidu}

\author{Hao Liu$^{\dagger}$, Qian Gao$^{\dagger}$, Jiang Li, Xiaochao Liao, Hao Xiong, Guangxing Chen, Wenlin Wang,\\ Guobao Yang, Zhiwei Zha, Daxiang Dong, Dejing Dou, Haoyi Xiong$^{*}$ }
\affiliation{
Baidu Inc., Beijing \country{China}\\
\{liuhao30, gaoqian05, lijiang01, liaoxiaochao, xionghao02, chenguangxing, wangwenlin,\\ yangguobao, zhazhiwei, dongdaxiang, doudejing, xionghaoyi\}@baidu.com
}

\thanks{$^{\dagger}$~Equal contribution.\\
    $^{*}$~Corresponding author.\\
    ``JiZhi'' means making smart decisions extremely fast in Chinese. An open-sourced version of \spred is available at https://github.com/PaddlePaddle/Serving
    }
\begin{abstract}
  In modern internet industries, deep learning based recommender systems have became an indispensable building block for a wide spectrum of applications, such as search engine, news feed, and short video clips.
  However, it remains challenging to carry the well-trained deep models for online real-time inference serving, with respect to the time-varying web-scale traffics from billions of users, in a cost-effective manner.
  In this work, we present \spred
  --- a Model-as-a-Service system --- that per second handles hundreds of millions of online inference requests to huge deep models with more than trillions of sparse parameters, for over twenty real-time recommendation services at Baidu, Inc.
  
  In \spred, the inference workflow of every recommendation request is transformed to a Staged Event-Driven Pipeline~(SEDP), where each node in the pipeline refers to a staged computation or I/O intensive task processor.
  With traffics of real-time inference requests arrived, each modularized processor can be run in a fully asynchronized way and managed separately. 
  Besides, \spred introduces the heterogeneous and hierarchical storage to further accelerate the online inference process by reducing unnecessary computations and potential data access latency induced by ultra-sparse model parameters.
  Moreover, an intelligent resource manager has been deployed to maximize the throughput of \spred over the shared infrastructure by searching the optimal resource allocation plan from historical logs and fine-tuning the load shedding policies over intermediate system feedback.
  Extensive experiments have been done to demonstrate the advantages of \spred from the perspectives of end-to-end service latency, system-wide throughput, and resource consumption.
  Since launched in July 2019, \spred has helped Baidu saved more than ten million US dollars in hardware and utility costs per year while handling 200\% more traffics without sacrificing the inference efficiency.
\end{abstract}


\maketitle

{\fontsize{8pt}{8pt} \selectfont
\textbf{ACM Reference Format:}\\
Hao Liu, Qian Gao, Jiang Li, Xiaochao Liao, Hao Xiong, Guangxing Chen, Wenlin Wang, Guobao Yang, Zhiwei Zha, Daxiang Dong, Dejing Dou, Haoyi Xiong. 2021. \spred: A Fast and Cost-Effective Model-As-A-Service System for Web-Scale Online Inference at Baidu. In \textit{Proceedings of the 27th International ACM SIGKDD Conference on Knowledge Discovery and Data Mining (KDD’21), August 14–18, 2021, Virtual Event, Singapore.} ACM, New York, NY, USA, 9 pages. https://doi.org/10.1145/3447548.3467146}

\section{Introduction}\label{sec:intro}
\begin{figure}[t]
	\begin{minipage}{1.0\linewidth}
		\centering
		\subfigure[{\small Baidu news feed.}]{\label{fig:newsfeed}
        \includegraphics[width=0.315\textwidth]{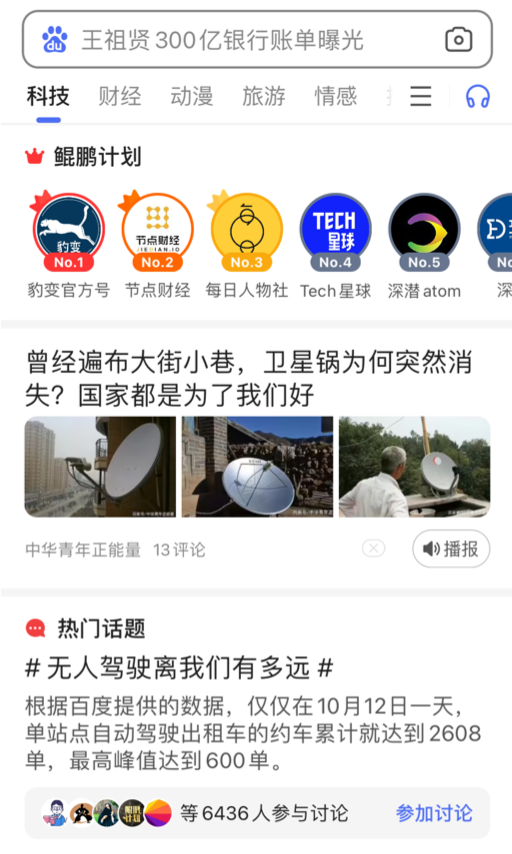}}
        \subfigure[{\small Haokan video.}]{\label{fig:haokan}
        \includegraphics[width=0.315\textwidth]{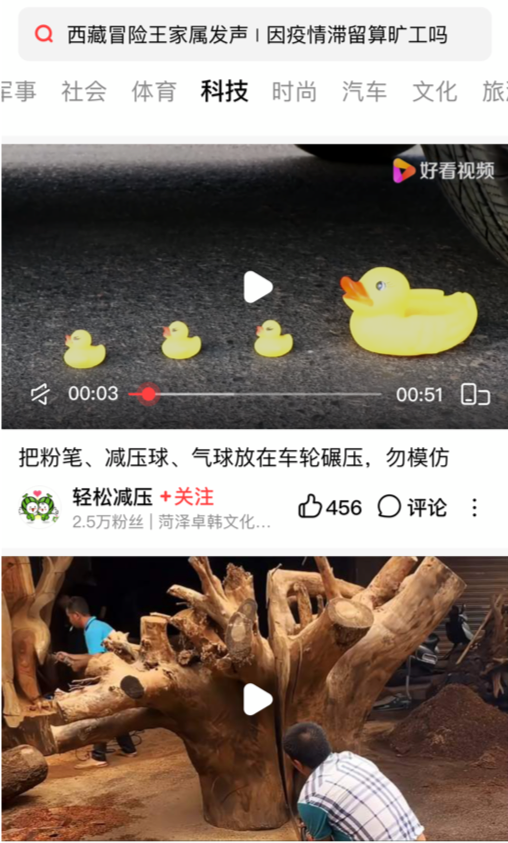}}
        \subfigure[{\small Baidu tieba.}]{\label{fig:haokan}
        \includegraphics[width=0.315\textwidth]{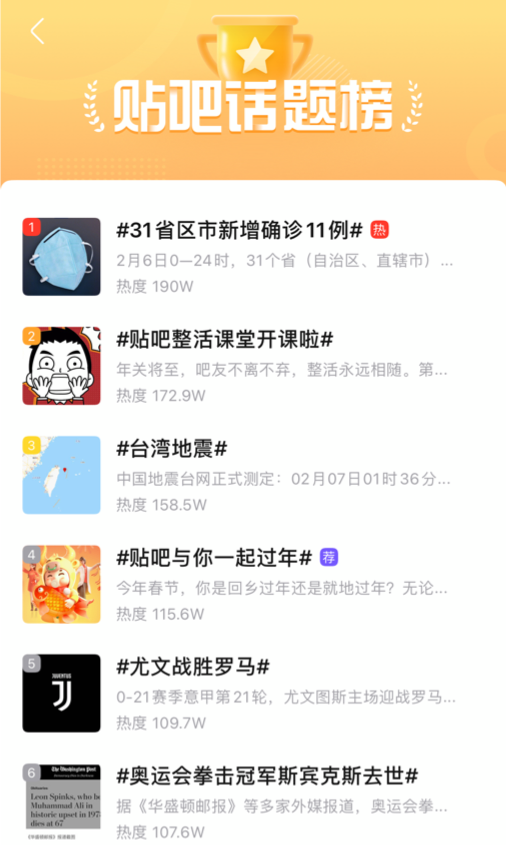}}
	\end{minipage}
	\vspace{-12pt}
	\caption{Screenshots of several DNN empowered products in Baidu based on \spred.}
    \vspace{-15pt} 
    \label{fig:product}
\end{figure}

Deep neural networks (DNNs)~\cite{lecun2015deep} have recently been used as the standard workhorses for predicting users' online behaviors, such as click-through rate~\cite{zhou2018deep,fu2020deep}, browsing interests~\cite{yang2019aiads}, and buying intentions~\cite{guo2019buying}, due to its superiority of representation modeling and reasoning for a wide spectrum of data modalities~\cite{ngiam2011multimodal}.
As early as in the early 2010s, leading internet companies such as Baidu and Google have successfully integrated DNNs into their core products, ranging from sponsored search~\cite{ouyang2014sda,fan2019mobius} to voice assistant~\cite{mohamed2011acoustic}.
\figref{fig:product} depicts several real-world products at Baidu that leveraging DNNs for online recommendations.

While DNNs have demonstrated its effectiveness in various internet application domains, the cost of using DNNs for web-scale real-time online inference becomes the major burden for most companies to adopt the techniques~\cite{jouppi2017datacenter,fowers2018configurable}
On the one hand, the time consumption (\eg latency) of the online service is critical for user experience~\cite{brutlag2009speed} and can influence the long term retention rate~\cite{bernardi2019150}.
On the other hand, the resource consumption (\eg hardware and energy usages) of supporting DNNs would request significant serving infrastructure investment~(\eg high-performance clusters) with higher power consumption and sometimes makes the systems design, implementation and operation over-budget~\cite{park2018deep}.

To guarantee the utility of online DNN serving, some efforts have been done from algorithmic and system perspectives. From the algorithmic perspective, model compression~\cite{cheng2017survey}
can significantly reduce the computational overhead by trade-off the model size and recommendation effectiveness. 
From the system perspective, dedicated acceleration hardware such as Graphics Processing Unit (GPU)~\cite{owens2008gpu}
and software extensions such as TensorRT~\cite{vanholder2016efficient} have been devised to reduce the DNN computation latency.
However, serving DNNs for online recommendations is more than executing the feed-forward neural network.
The aforementioned techniques are mainly optimized for single DNN training and inference, but overlook the complicated data and computation dependency of online inference under time-varying internet traffics. 

\begin{figure}[t]
	\begin{minipage}{1.0\linewidth}
		\centering
		\subfigure[{\small Traffic load and budget explosion.}]{\label{fig:traffic}
        \includegraphics[width=0.48\textwidth]{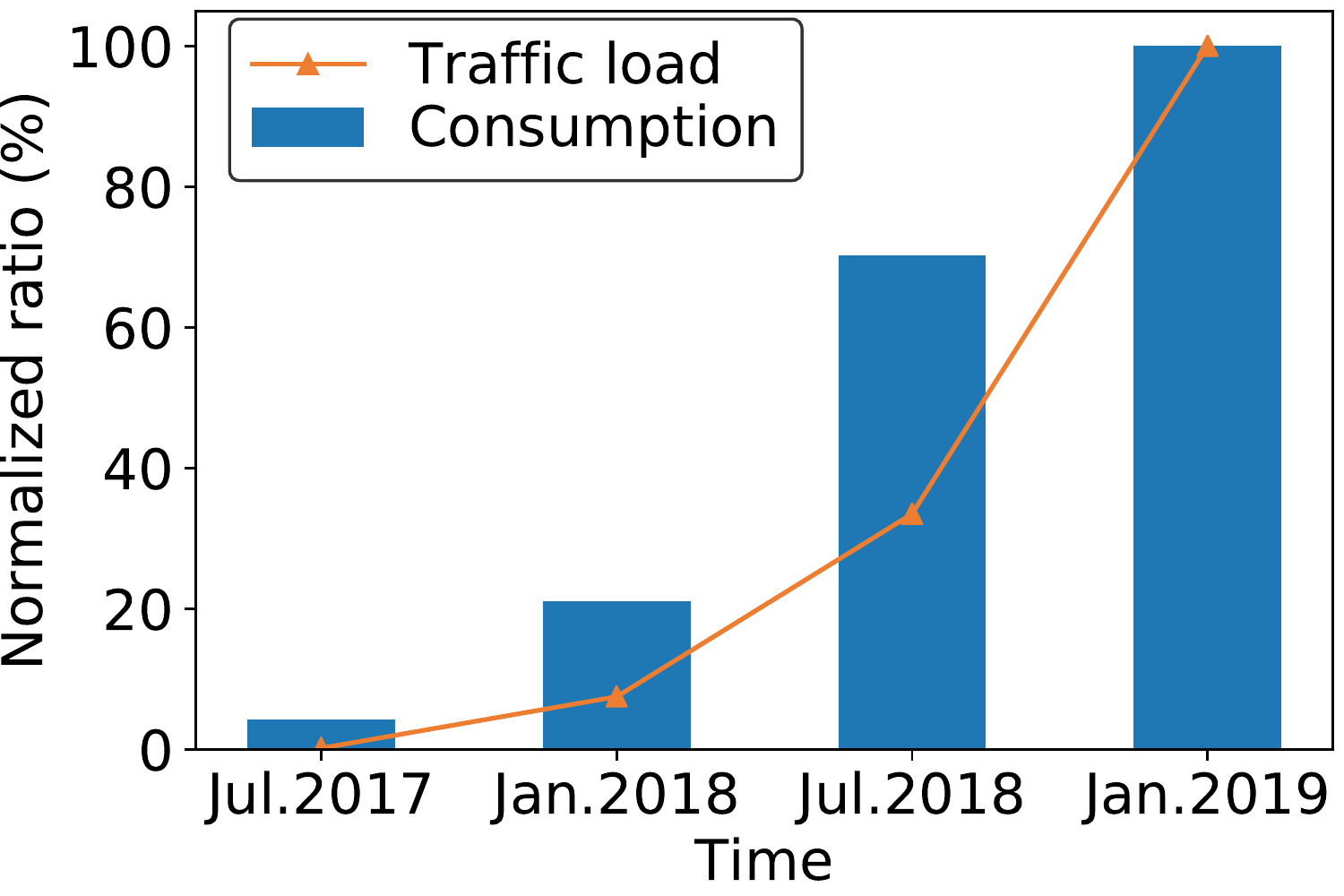}}
        \subfigure[{\small Online inference time breakdown.}]{\label{fig:breakdown}
        \includegraphics[width=0.48\textwidth]{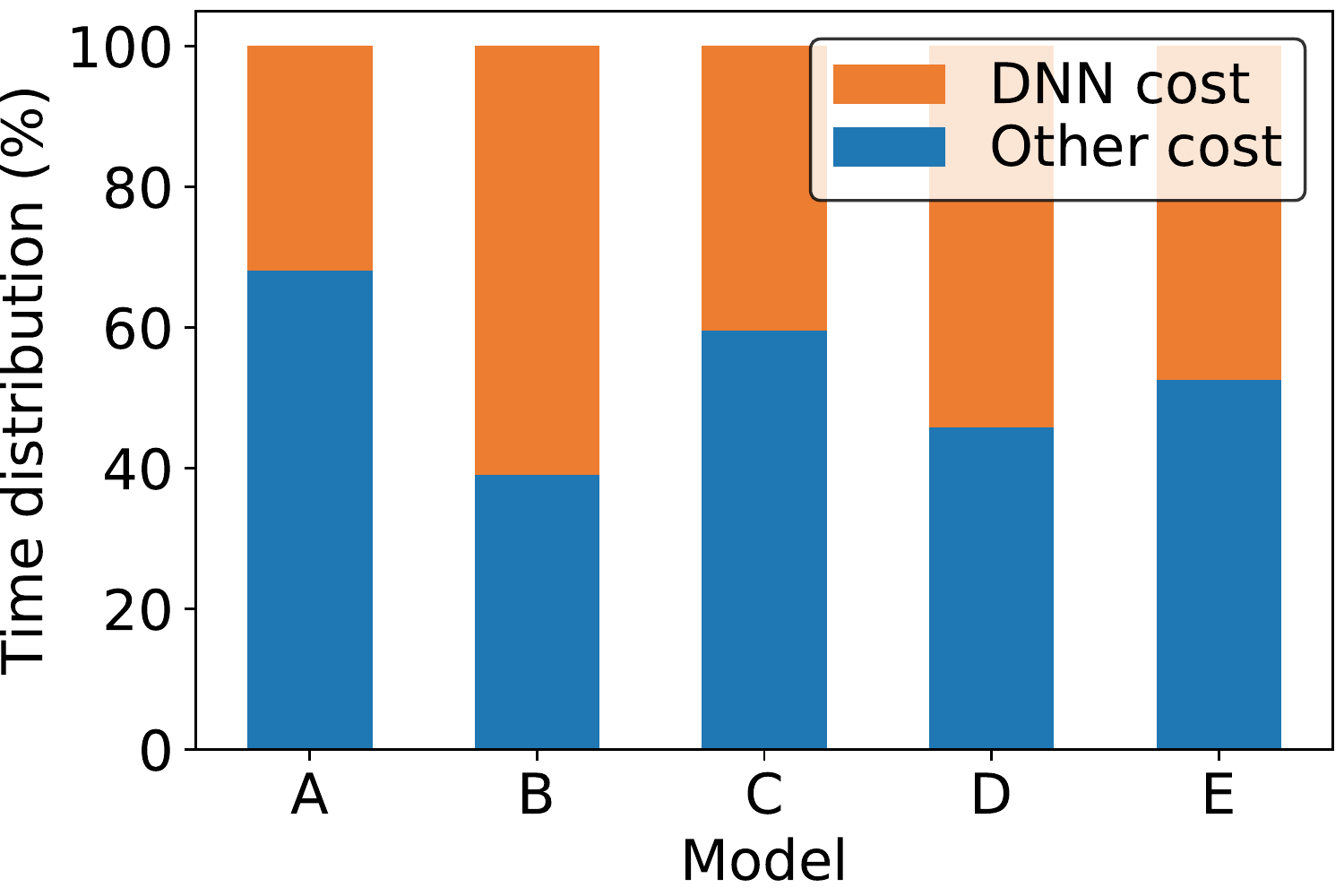}}
	\end{minipage}
	\vspace{-12pt}
	\caption{Characteristics of serving online recommendations. In past three years, Baidu expanded 200-fold budget on inference infrastructure to handle the explosive and diversified recommendation traffics.}
    \vspace{-15pt} 
    \label{fig:characteristics}
\end{figure}

Indeed, serving real-time DNN inference requests for web-scale recommendations is non-trivial for any major internet player with hundreds of millions of Daily Active Users (DAUs), because of the following three major technical challenges.
(1)~\emph{Huge and sparse DNN models}. To improve the recommendation performance, the industrial DNNs are usually trained with massive data samples, where each data sample is typically extremely high-dimensional and sparse (\ie hundreds billions of features with only hundreds of non-zero values). 
For example, the CTR model in Baidu sponsored search and Ads retrieval are over 10TB large~\cite{zhao2019aibox} and are still quickly growing.
As a result, these huge DNN models with trillions of sparse parameters require an efficient way to store, retrieve, transfer, and compute.
(2)~\emph{Time-varying web-scale traffics}.
The serving system of such deep learning based recommenders is responding to hundreds of millions concurrent inference requests per second.
More critically, the online inference service may experience 100-fold traffic increase in peak hour and fast traffic load growth, as depicted in \figref{fig:traffic}.
\eat{(\eg Baidu experienced 2,500 times higher load in the Spring Festival Gala red-envelope event in 2019~\cite{redenvelope})}
Since it is unrealistic to increase the budget for every single recommendation service unlimitedly, it is challenging to build an efficient and cost-effective online inference system for such explosive and temporally imbalanced traffic loads.
(3)~\emph{Diverse recommendation scenarios}.
\figref{fig:breakdown} shows the diversified data-processing and inference time distribution of five real-world deep learning based recommendation services at Baidu.
Depending on different application scenarios,
the recommendation strategy and inference pipeline may vary.
For instance, news recommendation usually handles both text and image related features and may involve multiple DNNs~(up to 20 DNNs in single recommendation service in Baidu), while search engine pays more attention to user preference modeling~\cite{yang2019aiads}.
It is challenging to build a unified online serving architecture that can support various recommendation scenarios with different online inference workflow.

In this work, we present \spred, the Model-as-a-Service system for web-scale online inference at Baidu. 
Specifically, \spred first transforms the workflow of deep learning based recommendation inference into the \emph{Staged Event-Driven Pipeline}~(SEDP), 
where each node refers to a staged computation task and the edge between two nodes refers to a possible transition from one task to another.
By modularizing the complex and application specific inference pipeline into individual staged processors, SEDP supports flexible customization, performance tuning, and can be well adapted for diverse application scenarios.
Moreover, a tailor-designed \emph{Heterogeneous and Hierarchical Storage}~(HHS) module is introduced for the huge and ultra-sparse DNN model management. By eliminating redundant computations and parameter access costs, HHS significantly improves the online inference speed and throughput in a cost-effective way.
Besides, to improve the computational resource utilization, we propose the \emph{Intelligent Resource Manager}~(IRM) module for automatic system performance tuning. By formulating auto-tuning tasks as constrained optimization problems, IRM searches the optimal resource allocation plan for each stage processor in the offline and adaptively controls the online load shedding policy based on real-time system feedback, without sacrificing the recommendation effectiveness.

To the best of our knowledge, this work is the first to study the system design, implementation, and integration for web-scale DNN inference from an industrial practitioner's perspective, by addressing the goals of minimizing the overall resource consumption under latency constraints. 
Our major contributions are summarized as follows.
(1)~We propose an asynchronous design of online inference for deep learning based recommendation, which exposes the opportunity of fine-grained resource management and flexible workflow customization.
(2)~We introduce a cost-effective data placement strategy for huge and ultra-sparse DNN models, which significantly improves the online inference system throughput with slight resource overhead.
(3)~We propose a novel intelligent resource management module to optimize the system resource utilization in both online and offline fashion.
(4)~We conduct extensive experiments to demonstrate the advantages of \spred by using real-world web-scale traffics from various products in Baidu. Moreover, we share our hands on experiences of implementing and deploying the proposed system, which we envision would be helpful to the community.
We have released an open-source version of \spred on PaddlePaddle (https://github.com/PaddlePaddle/Serving).

\section{Background and Related Work}\label{sec:back}
This section overviews the common routine of industrial deep learning based online recommendation and online inference speedup techniques, including the legacy online serving design at Baidu.

\textbf{Deep learning based online recommendation}.
Given a set of items~(\eg contents and goods), the recommendation system aims to return personalized probability score for each item to achieve one or more targets, such as maximizing the CTR~\cite{zhou2018deep}, improving the dwell time~\cite{lamba2019modeling}, and increase the overall profitability of the product~\cite{yang2019aiads}.
In the past years, DNN has been widely adopted to serve as the core model in online recommendation~\cite{hu2018web,liu2020incorporating,liu2021vldb}.
Based on the application scenario, different DNN architectures have been adopted for recommendation. For instance, neural collaborative filtering for user-item interaction modeling~\cite{he2017neural}, recurrent neural network for sequential user behavior modeling~\cite{hidasi2015session,qin2020enhanced}, and deep reinforcement learning for long-term recommendation optimization~\cite{zou2019reinforcement}.
In Baidu, DNN based recommendation service has been successfully integrated into over twenty online internet products (\eg news feed, short video clips, search engine, and online advertising), serving over one hundred billion online recommendation requests made by hundreds of millions users per day.

\textbf{Serving deep learning based online inference}.
To guarantee the user experience, the online inference process is usually constrained to be finished in an ultra low latency, \ie dozens of milliseconds.
However, the majority of research efforts on DNN speedup are focused on model training~\cite{zhang2013asynchronous,chen2014dadiannao}, while limited efforts have been made for online inference.
On the one hand, various model compression techniques~\cite{cheng2017survey} have been proposed to reduce the model size by sacrificing the recommendation effectiveness.
On the other hand, dedicated processors~(\eg GPU~\cite{owens2008gpu}, TPU~\cite{jouppi2017datacenter}, Kunlun~\cite{ouyang2020baidu}) and software extensions~(\eg AVX\cite{firasta2008intel}, SSE~\cite{amiri2020simd}, TensorRT~\cite{vanholder2016efficient}, and TensorFlow-Serving~\cite{46801}) have been developed to accelerate online inference computation.
However, the above approaches are mainly optimized for general DNN acceleration, but overlook the unique challenge of diversified data dependency and model sparsity in the online recommendation.

\textbf{The previous system design for online inference at Baidu}.
Since 2013, Baidu has proposed dedicated hardware and software platforms to accelerate online data processing and DNN inference~\cite{zhao2019aibox,ouyang2020baidu}.
When real-time recommendation requests arrive, the system first retrieves relevant candidate items (usually millions) and then apply multi-phase inference for user-item pairs~\cite{fan2019mobius} to reduce the candidate size from millions to thousands and finally return a dozen results to the user.
In each phase, by following the data parallel paradigm, the inference tasks~(user-item pairs) are packaged into multiple batches, and different batches are then processed in parallel.
For each inference task, the online inference system requires to fetch corresponding raw data from remotely distributed storage, transform input to proper features, and execute the feed-forward inference computation to output the relevance score.
For different recommendation applications, the online inference service needs to be developed from scratch and tuning based on expert experience.
More critically, the system suffers from pipeline stall because of the computation and access latency of long-tail candidates,
and quickly meets the resource bottleneck under the explosive recommendation traffic demand.

\section{System Overview}\label{sec:system}
\begin{figure}[t]
    \centering
    \includegraphics[width=.45\textwidth]{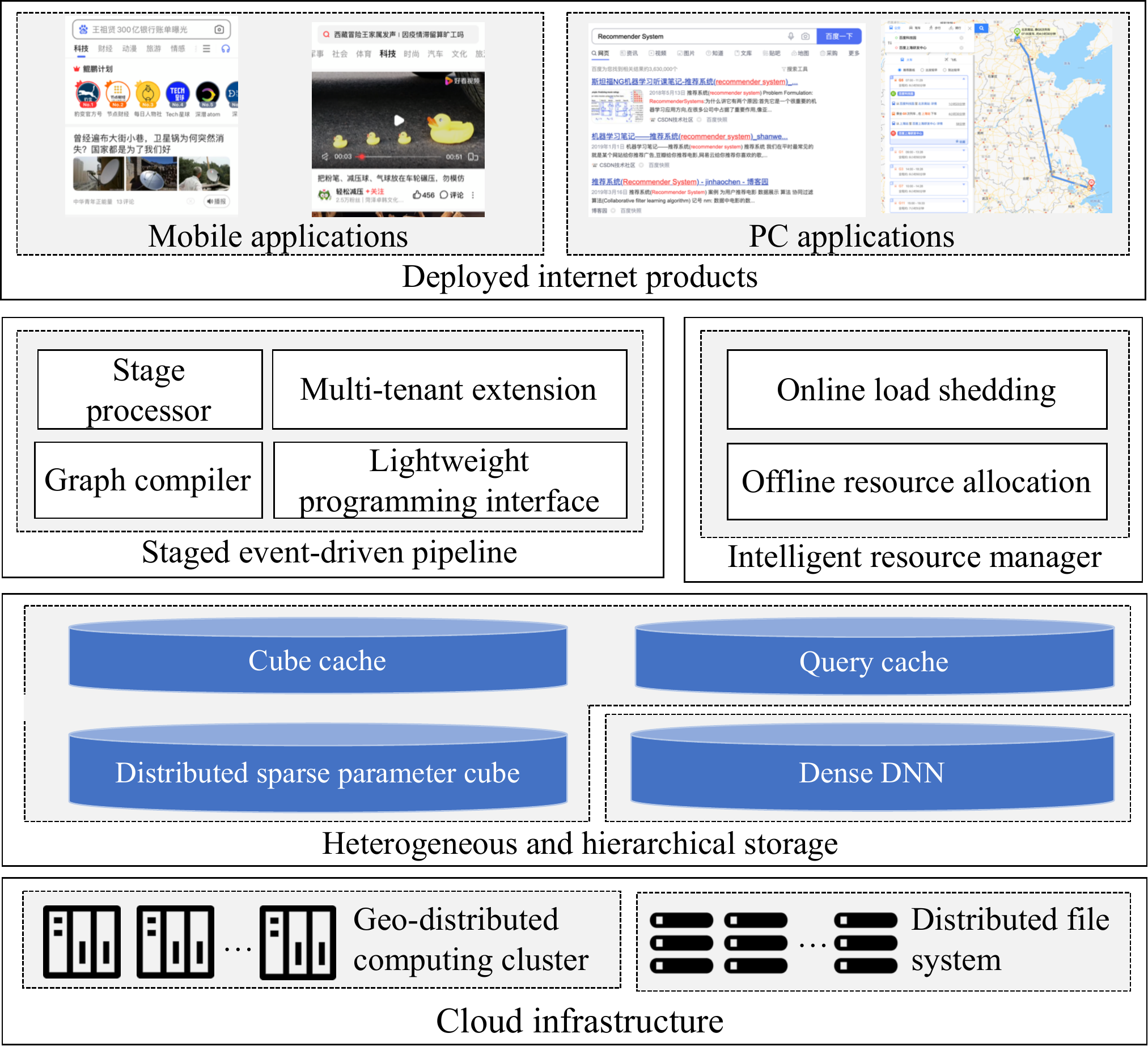}
    \vspace{-10pt}
    \caption{Layered architecture of \spred.}
    \vspace{-15pt}
    \label{fig:arctitecture}
\end{figure}
\figref{fig:arctitecture} shows the architecture of \spred, which consists of three major components, the \emph{Staged event-driven pipeline} (SEDP), the \emph{Heterogeneous and hierarchical storage}, and the \emph{Intelligent resource manager}.
Given an online inference task, the \emph{Staged event-driven pipeline} first compiles the inference workflow into a directed acyclic graph and then execute each independent node in the graph in a full asynchronous way.
By modularizing the inference task into a set of sequence or parallel stages, SEDP can significantly improve the resource utilization of the infrastructure, enable workflow customization, and expose opportunities for more intelligent dynamic resource allocation.
The \emph{Heterogeneous and Hierarchical Storage}~(HHS) is a key-value based data placement system including a distributed sparse parameter cube and a heterogeneous and hierarchical cache.
By eliminating unnecessary computation and various I/O costs, the storage system further reduces the online inference latency and improves the system throughput in a cost-effective way.
Based on SEDP, the \emph{Intelligent Resource Manager}~(IRM) achieves quasi-optimal resource utilization by monitoring and automatically tuning the computation resource allocation of the serving pipeline in both offline and online. 
In particular, the offline resource management component searches the optimal resource allocation plan for each stage in the SEDP and HHS to maximize the system throughput. 
The online resource management module adaptively fine-tunes the load shedding policies based on the intermediate system feedback to guarantee the serving latency when the traffic is overloaded, without sacrificing the recommendation effectiveness. 

\begin{figure*}[t]
    \centering
    \includegraphics[width=.8\textwidth]{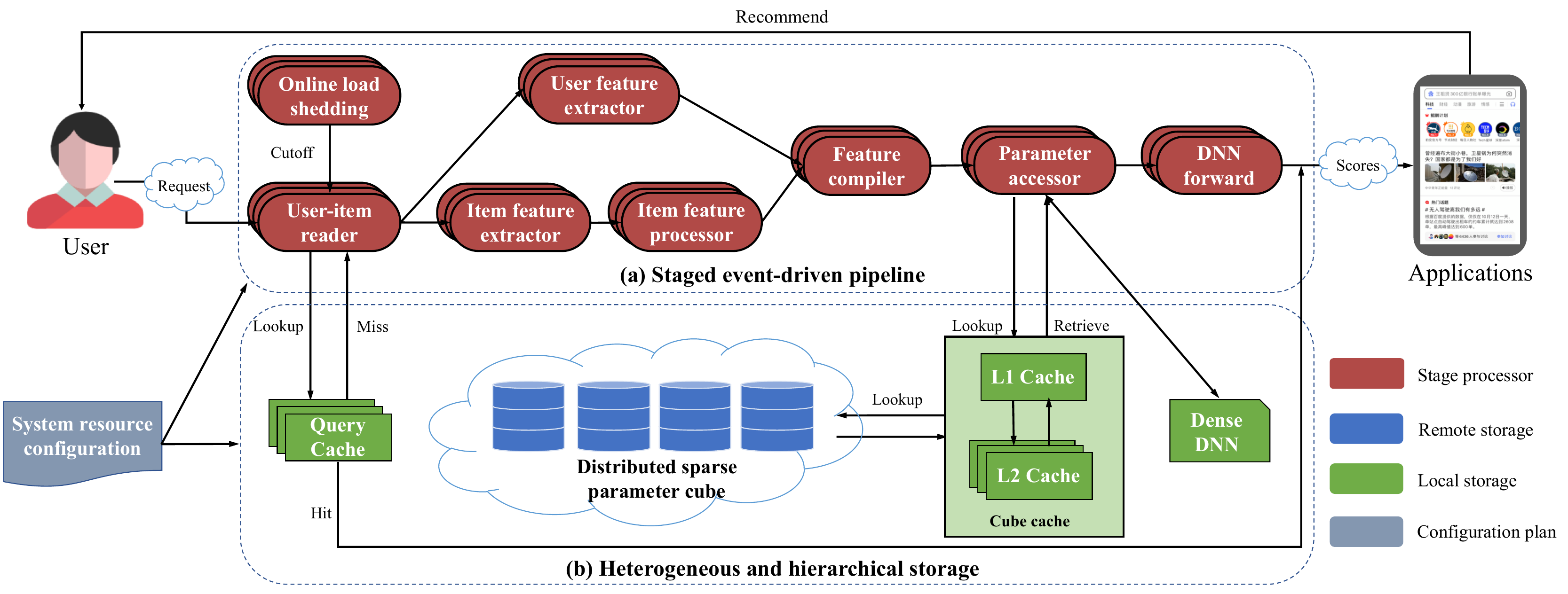}
    \vspace{-10pt}
    \caption{An illustrative online user-item inference workflow based on \spred.}
    \vspace{-8pt}
    \label{fig:workflow}
\end{figure*}

\section{Staged Event-Driven Pipeline}
The primary design objective of \spred is to provide customizable and tunable online inference service for diverse online recommendation scenarios, which is achieved by the asynchronous \emph{Staged Event-Driven Pipeline}.
We begin with the construction of stage processor.

\begin{defi}
\textbf{Stage processor}. Given an online inference workflow, a stage processor $p$ is defined as a tuple $p=\langle op, c\rangle$, where $op$ is the operator representing a unit primitive that encapsulate the functionality for a particular execution stage, and $c$ is the channel that queuing events from upstream processors via a shared data structure.
\end{defi}

Note an execution stage can be any steps in the online inference workflow, such as feature extraction, model access, and DNN feed-forward propagation.
In this way, an online inference workflow is modularized into a set of stage processors $P=\{p_1, p_2, \dots\}$.
A stage processor can be understood as a micro-service or a thread, which is a reusable and self-contained component that may appear one or multiple times in a workflow depend on the online inference logic.
In each stage processor, the operator processes incoming events concurrently, and passes the finished events to the next channel queue attached to subsequent stage processors.

\begin{defi}
    \textbf{Staged Event-Driven Pipeline (SEDP)}. A staged event-driven pipeline is defined as a directed acyclic graph $G=(P,E)$, where $p_i \in P$ is a stage operator, and $e_{ij} \in E$ is an edge connecting two stage processors $p_i$ and $p_j$. Note that in a SEDP, all edges point to processors of a same stage share a same channel, to provide complex event processing capability, such as aggregation, join, and ordering.
\end{defi}

For a predefined SEDP, \spred can automatically analyze the dependency between each stage processor, construct shared channels, and compile the fully asynchronous execution plan.
In particular, each arrived online inference request (\ie user-item pair) is regarded as an event and buffered in a queue of the next stage processor.
Based on the graph dependency, each stage processor in a SEDP can be executed in sequence or parallel.

\figref{fig:workflow}~(a) depicts an example online workflow of user-item inference task by using SEDP.
The user and item information is first collected and dispatched to a user-specific processor and item-specific processors. After internal processing stages such as feature filtering, ranking, and wrangling, all raw features of the request are recombined to retrieve the corresponding feature parameters~(\ie pre-computed embeddings and sparse vectors). Finally, the retrieved sparse parameters are fed to the DNN network to derive the item relevance score for personalized recommendation.

By asynchronously executing stage processors in an event-driven way, SEDP successfully improves the overall throughput of the online inference system by reducing potential pipeline stalls of long-tail items. 
Note that the capacity~(\eg parallelism, memory) of each stage processor can be tuned separately to achieve a higher resource utilization rate of the shared infrastructure.
Moreover, the modularized design exposes the customizable online inference workflow interface to engineers and data scientists to support fast service workflow and model iteration.

\textbf{Multi-tenant extension}.
SEDP also offers the flexibility to simultaneously access multiple DNNs in a single pipeline, \eg multi-phase inference where each phase corresponds to a DNN, and multi-objective inference accesses multiple DNNs in parallel.
The multi-tenant extension not only enables more complex recommendation strategies in a unified service, but also benefits the recommendation system iteration via the natural support of online A/B testing.
Specifically, the access of multiple DNNs usually involves tremendous repetitive data processing and sparse parameter accessing operations, which can be significantly reduced in a single service.
Moreover, in industrial product iteration, data scientists and researchers often conduct tremendous A/B testing daily~(usually tens of test groups in concurrent).
A common approach is to conduct A/B testing between the current recommendation system and a new test group to determine the satisfaction of the new model/strategy.
In the previous online inference system, the system administrator needs to deploy new online services for each new test group, which requires manually splitting the traffic load and assigning computational resources for different variants.
However, such an approach is cumbersome and error-prone.
In \spred, the A/B testing is well-supported by hosting multiple different test groups in a multi-tenant way.
By devising a stage processor to dispatch traffic loads to different test groups, each variant is an independent branch in the SEDP and shares the same computing infrastructure.  

\section{Heterogeneous and Hierarchical Storage}
We further present the cost-effective data placement design, including the distributed sparse parameter cube and the heterogeneous and hierarchical cache.
Both components are tailor-designed for deep learning based recommendation in a data-driven approach and tunable to handle different recommendation scenarios.

\subsection{The Distributed Sparse Parameter Cube}
In online recommendation, the DNN usually consists of a huge and sparse sub-network~(up to Terabytes large) and a moderate-sized dense network~(usually hundreds of Megabytes)~\cite{zhao2019aibox}.
We first introduce the distributed sparse parameter cube (\aka the cube) to manage the extremely high-dimensional parameters of the sparse sub-network. 
Specifically, the cube is a read-only distributed key-value store, where the key is defined as a compact feature signature~(a universally unique identifier of a feature), and the value is defined as model weights and statistics of user feedback of the corresponding sparse feature.
To guarantee the distributed consistency, the feature signature is derived via universal hash function~\cite{carter1979universal}.
The online inference tasks can look up the extremely sparse non-zero values in constant time complexity by accessing the cube.
To hide the latency induced by random access of the hash table, all keys are stored purely in memory, and values are grouped into one Gigabyte-sized block that can be either placed in memory or disk~(\ie SSD).
The placement of keys and values is a trade-off between the online inference latency and the hardware consumption.
In this way, the cube operates on heterogeneous hardware can support ten-thousand level concurrent parameter lookup with less than ten milliseconds latency within a data center.
The cube is also fault-tolerant by replicating data blocks in the distributed cluster.

\subsection{Heterogeneous and Hierarchical Cache}
Although the distributed sparse parameter cube successfully reduces the feature access latency to several milliseconds, the online inference process is still communication and computational extensive.
We further introduce the heterogeneous and hierarchical cache to reduce both the I/O and computation costs, including the cube cache for the distributed sparse parameter cube access and the query cache for the inference pipeline computation.
\figref{fig:workflow}~(b) shows two types of caches and their interaction with the SEDP.

\begin{figure}[t]
	\begin{minipage}{1.0\linewidth}
    \vspace{-2pt} 
		\centering
		\subfigure[{\small Access distribution of sparse parameters.}]{\label{fig:cube-pattern}
        \includegraphics[width=0.48\textwidth]{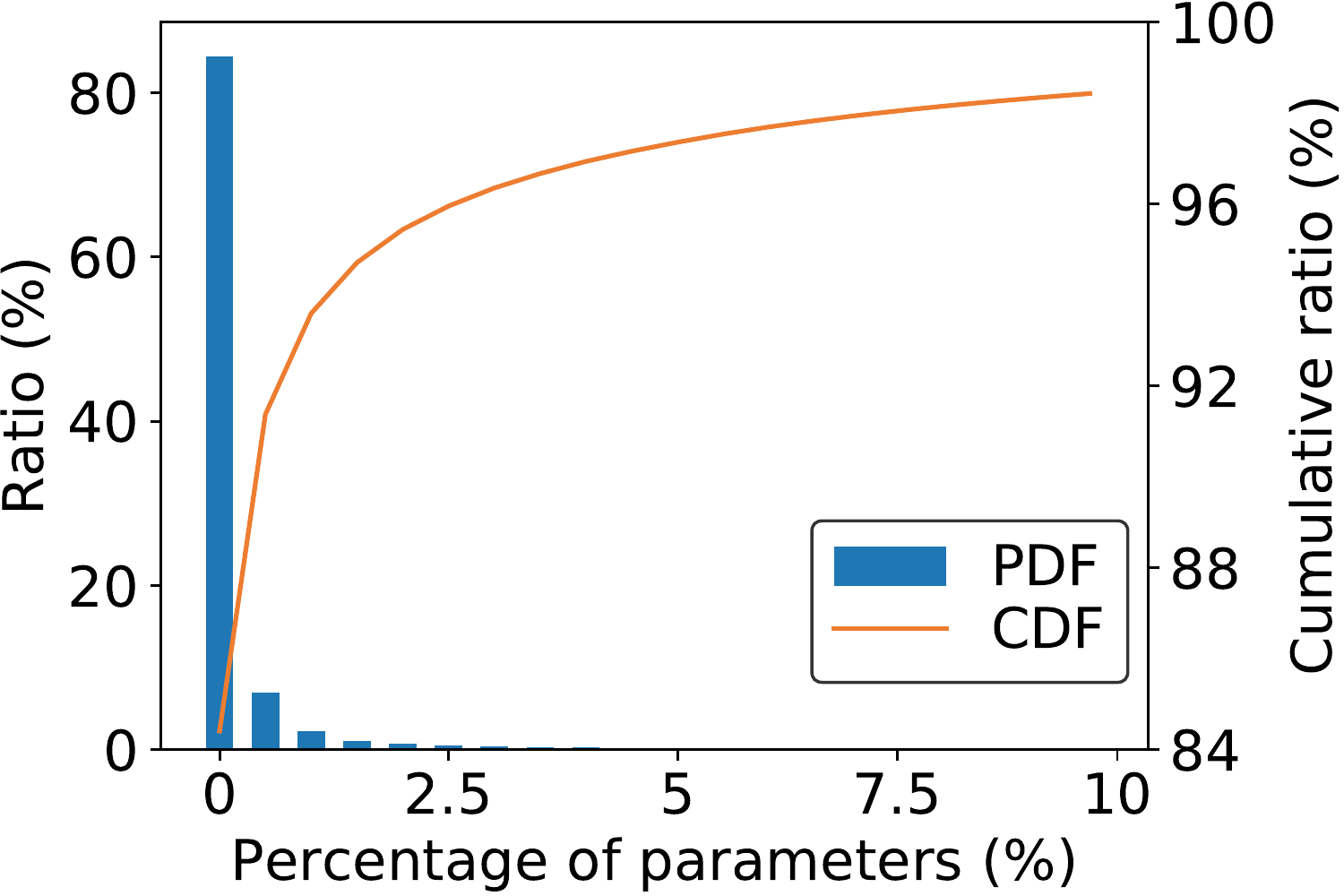}}
        \subfigure[{\small Recurrence distribution and variance of user-item pairs.}]{\label{fig:q-pattern}
        \includegraphics[width=0.48\textwidth]{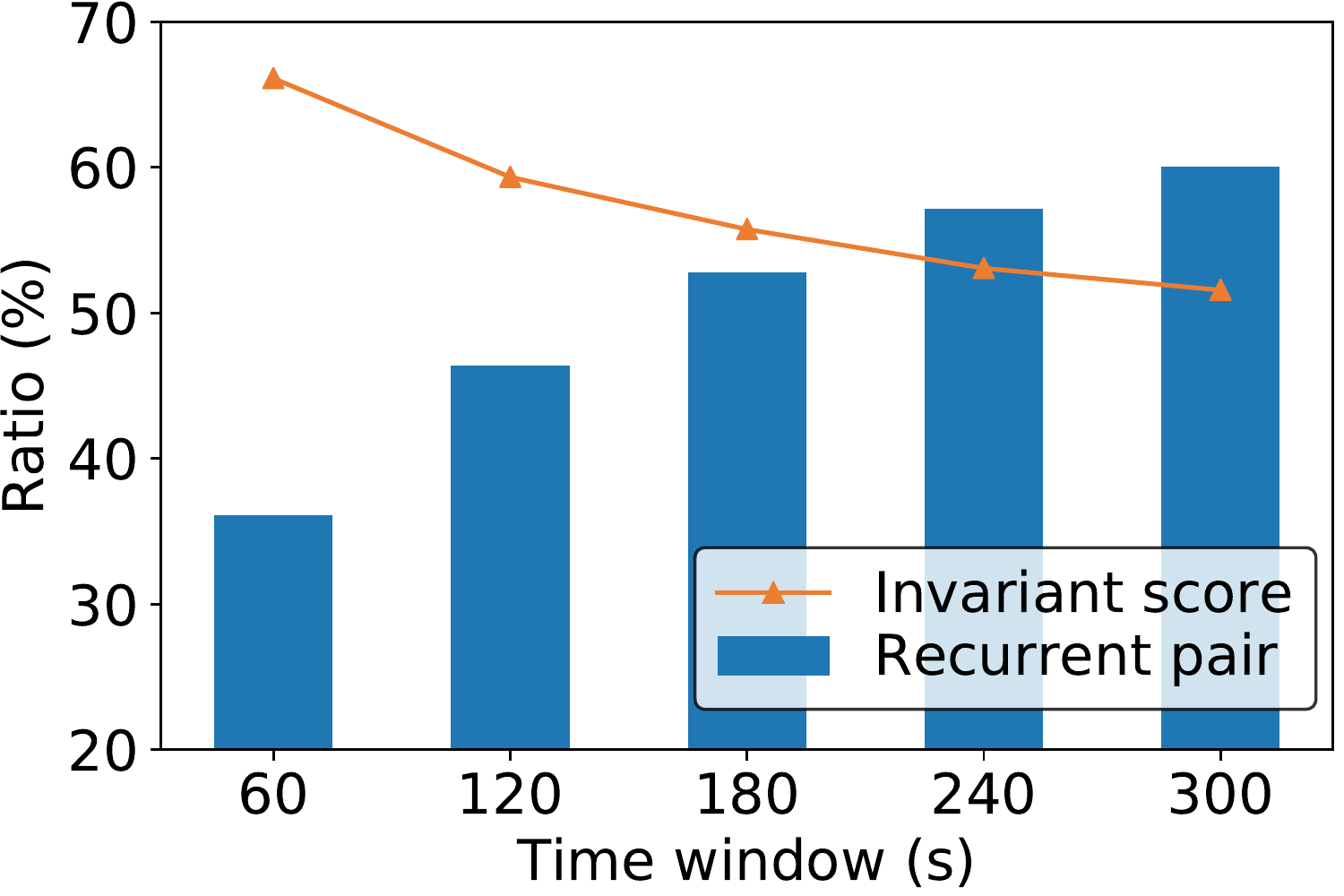}}
	\end{minipage}
	\vspace{-10pt}
	\caption{Data access patterns in online inference service.}
    \vspace{-15pt} 
    \label{fig:data-pattern}
\end{figure}

\textbf{Cube cache}. The design of the cube cache is motivated by the observation that the feature access to the cube is heavy-tailed, where only a small portion of hot features are frequently accessed and updated.
\figref{fig:cube-pattern} plots the distribution of 100 million consecutive cube accesses of a deployed online recommendation service in Baidu, where over $80\%$ lookup are focused on $1\%$ values in the cube.
Therefore, it is reasonable to cache such hot features locally to reduce expensive network I/O costs.
Specifically, the cube cache is a two-layer local storage that persistent a small portion of frequently accessed key-value pairs in the cube.
We adopt the Least Frequently Used~(LFU)~\cite{lee1999existence} policy to replace key-value pairs in the cache.
There are two levels of the cube cache, \ie the memory-level and the disk-level.
The disk-level cache is used to hide the expensive network I/O cost of accessing the remote cube, which stores about $1\%$ frequently accessed key-value pairs to local SSDs.
On top of the disk cache, a memory cache is further introduced to avoid disk I/O for the top $0.1\%$ hottest key-value pairs.
Thanks to the cube cache, the SEDP can avoid up to $90\%$ remote accesses, which reduces $10\%$ latency on average.

\textbf{Query cache}. 
We further introduce the query cache to eliminate unnecessary computations in online inference.
Remember the general target of the online recommendation service is to derive the user-item relevance score.
The key design insight of the query cache is that the user-item score is stable in a short time period~(\eg several minutes) under certain conditions.
\figref{fig:q-pattern} depicts the timeliness of the calculated user-item score. As can be seen, by bounding the time window to two minutes, over $60\%$ scores are invariant.
Therefore, for certain recently processed user-item pairs, the computation can be safely eliminated by reusing the previously calculated score, with little influence on the overall recommendation performance.
The query cache is designed as purely in-memory storage, which persists a set of recently calculated user-item scores that fulfill certain conditions~(\eg high relevance items).
To guarantee the recommendation effectiveness, the cached scores will expire after a tunable time window or any user feedback~(\eg click, unlike), which indicates the intermediate change of user preference.
Unlike the cube cache, the query cache employs the Least Recently Used~(LRU) policy.
Query cache helps \spred reduces over $20\%$ computations, which significantly improves the cost-effectiveness of the online inference service.

\section{Intelligent resource manager}
In this section, we describes the \emph{Intelligent Resource Manager}~(IRM) in detail, including both offline and online auto-tuning components.
By automatically searching and learning the resource allocation and the load shedding policies for SEDP and HHS, IRM shields engineers and data scientists from the complexity of performance tuning, and achieves even better performance than expert prior.

\subsection{Offline Auto-Tuning for Quasi-Optimal Resource Allocation}
Beyond tuning every single stage processors and caches independently, \spred optimizes the parameters of the system on a shared computing infrastructure in a global manner.
Specifically, the goal of offline optimization is to minimize the resource consumption under the latency constraint.
We first categorize the tunable system parameters into two classes. (1)~\emph{System level} parameters that influence the global system.
(2)~\emph{Stage level} parameters that work on dedicated stage processors.
Please refer to Appendix~\ref{appendix:extend-param} for more detailed offline optimization parameters. 

Specifically, based on historical logs of every stage processor with the varying parameters, \spred collectively searches the global optimal resource allocation plan.
The optimization objective is 
\begin{equation}\label{eq:opt}
    \begin{aligned}
        \mathcal{O}_\mathrm{offline}  & =   \underset{(\Theta,\theta_1,\theta_2,\dots,\theta_N)\in\Gamma}{\mathrm{arg min}}\ \sum_{j=1}^N\mathcal{F}^R_j(\Theta;\theta_j), \\
        & s.t.\ \mathcal{F}^L_j(\Theta,\theta_j)\leq \mathcal{F}^L_j(\bar\Theta,\bar\theta_j),\ \forall1\leq j\leq N,
    \end{aligned}
\end{equation}
where $N$ refers to the number of stage processors in the system, $\Gamma$ denotes the set of all possible parameters, $\Theta$ refers to the system level parameters, $\theta_j$ for $1\leq j\leq N$ refers to the stage level parameters for the $j^{th}$ stage processor, $\bar\Theta$ and $\bar\theta_j$ are the default settings of the parameters. The model $\mathcal{F}^R_j(\Theta;\theta_j)$ predicts the resource consumption of the $j^{th}$ stage processor based on the parameters of $\Theta$ and $\theta_j$, while the model $\mathcal{F}^L_j(\Theta;\theta_j)$ predicts the end-to-end latency of the $j^{th}$ stage processor based on the parameters of $\Theta$ and $\theta_j$. This optimization problem is challenging as (1) the feasible parameters should be able to bound the end-to-end latency of every stage processor (i.e., $N$ constraints), (2) $\mathcal{F}^R_j(\Theta;\theta_j)$ and $\mathcal{F}^L_j(\Theta;\theta_j)$ are built using ensemble of practical regression models, which are not naturally non-differentiable over the parameter spaces, and (3) the prediction results of $\mathcal{F}^R_j(\Theta;\theta_j)$ and $\mathcal{F}^L_j(\Theta;\theta_j)$ are noisy and probably biased compared to the ground truth. Thus, in addition to common gradient-based approaches, global optimization is required for parameter search.
\spred devises the Covariance Matrix Adaptation -- Evolution Strategy (CMA-ES) with constraints~\cite{arnold20121+} for optimal plan search. The objective and constraints are all encoded as black-box functions, while CMA-ES with constraints handles these functions naturally for global optimization in a sampling manner. Note that, in addition to using the minimizers obtained by CMA-ES, \spred restores the solution paths of CMA-ES (\ie the parameters having been ever reached), pickup constraint-satisfied results with the minimal objectives from solution paths, and evaluate these results in the realistic system deployment~(\ie using $5\%$ of overall online requests) to find the global optimal parameters. 

\subsection{Online Resource Management for Adaptive Load Shedding}
Different with the offline auto-tuning, the online resource management aims to handle the time-varying traffic load by swiftly deriving fine-grained computation policy for each online inference task based on real-time system feed-backs.
The online resource management is more tightly coupled with the recommendation service logic and usually requires task-specific optimization.
In this paper, we showcase a commonly used online load shedding module integrated in various online inference services in Baidu, to demonstrate the cost-effectiveness of online resource management.

As aforementioned, most industrial recommendation services adopt a funnel-shaped pipeline~\cite{fan2019mobius} by first retrieving a subset of potential candidates based on lightweight features~(\aka the recall phase) and then scoring each candidate more accurately based on more computation expensive features~(\aka re-ranking phase).
In most cases, only a dozen items will eventually be recommended to users, but over three orders of magnitude candidates will be passed to the computational expensive re-ranking stage processors.
The key insight of online load shedding is to adaptively prune low-quality candidates when the traffic load exceed the system capacity, with the minimum recommendation effectiveness degradation constraint.
Formally, given a set of selected candidate items $D^{inf}=\{d_1, d_2, $ $\dots, d_n\}$ from the recall phase, the objective of the online load shedding component is
\begin{equation}
\begin{aligned}
    \mathcal{O}_\mathrm{online} = &\arg\max_{q} \frac{1}{K}\sum^K_{i=1}\mathcal{C}(p_i(q_i \circ D_i^{inf})),\\
    &s.t.\ |\mathcal{L}^*-\hat{\mathcal{L}}| \leq \epsilon, 
\end{aligned}
\end{equation}
where $K$ is the total number of recommendation requests, $\mathcal{C}$ is the computational cost function, $p_i$ is the corresponding stage processor, $q_i$ is the pruning function conduct element-wise pruning on $D_i^{inf}$, $\mathcal{L}^*$ and $\hat{\mathcal{L}}$ are respectively the accuracy before and after online load shedding,
$\epsilon$ is a small value bounds the recommendation effectiveness degradation.

\begin{figure}[t]
    \centering
    \includegraphics[width=.4\textwidth]{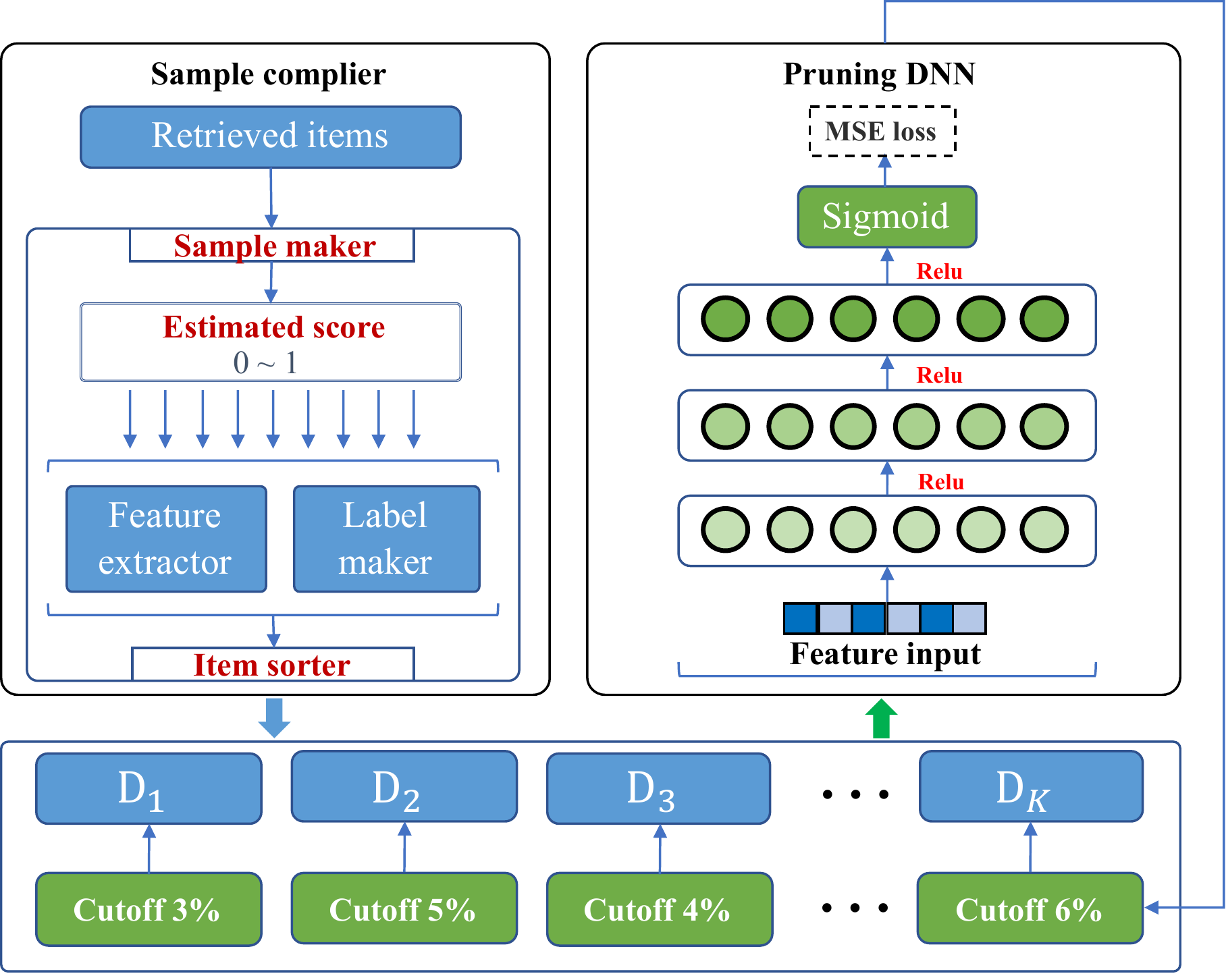}
    \vspace{-10pt}
    \caption{Pruning DNN based online load shedding.}
    \vspace{-15pt}
    \label{fig:onlinednn}
\end{figure}

To find the optimal pruning policy in an ultra-efficiency way, we transform the constrained optimization problem to a regression problem and employ a lightweight DNN~(\aka the pruning DNN) to decide the number of candidates to cutoff.
Specifically, we first sort $D_i^{inf}$ by leveraging the estimated score from the recall phase.
Then we apply the well-trained DNN to decide the cutoff line for each $D_i^{inf}$, where the candidates behind the cutoff line will be directly dropped without passing to the subsequent re-ranking stage processors.
\figref{fig:onlinednn} depicts the workflow of the online load shedding process.
Based on extracted features from the upstream stage processors and context features such as the intermediate system latency, the pruning DNN is ultra-lightweight and can prune low-quality candidates in dozens of microseconds. Please refer to Appendix~\ref{appendix:extend-param} for detailed feature list.
Different with general load shedding on data streams~\cite{tatbul2003load,babcock2007load}, the online load shedding in \spred adaptively decides the request-wise pruning ratio by taking the global recommendation effectiveness into consideration.

\section{Deployment} 
\spred has been deployed in the production environment to support a variety of real-world products in Baidu. 
In this section, we discuss the deployment details.
Please refer to Appendix~\ref{appendix:extend-impl} for details of implementation.

\textbf{Online deployment}.
\spred answers remote service access via BRPC~(https://github.com/brpc/brpc), a scalable Remote Procedure Call framework used throughout Baidu. The online inference service is duplicated in data centers distributed across China to reduce potential network latency.
Note in \spred, the response latency and resource consumption is a design trade-off, where allocate more sources can improve the system throughput and reduce latency, and developers can reduce the hardware consumption by relaxing the latency in a reasonable range.

\textbf{Model management and hot-loading}. 
In \spred, the sparse part of the DNN is managed by the distributed sparse parameter cube and the dense part is directly stored in memory.
In real-world recommendation applications, the model is updated frequently~(\eg per hour), and it is undesired to update the model by frequently restart the service.
To guarantee the availability of the online service, \spred supports model hot loading without service interruption.
Specifically, we build a monitoring module to continuously track model update state in the remote address~(\ie the model training cluster).
Once a new model is ready~(identified by model generation timestamp), the new model will be pulled to local and loaded to serve new coming recommendation requests via double buffering.

\section{Experimental evaluation}\label{sec:exp}

\subsection{Experimental Setup}
In this section, we conduct extensive experiments to evaluate the effectiveness of the proposed system.
We report evaluations on four different \spred empowered real-world online inference services deployed in Baidu. All results are collected from the online living system.
Specifically, \emph{Service A} is for a multi-modal recommendation model on Baidu news feed, \emph{Service B} and \emph{Service C} are video recommendation services used in different product scenarios, \emph{Service D} is a multi-objective recommendation service fulfilling multiple targets.
\tabref{table-data} summarizes model statistics of five recommendation services.
We use the previous online inference system~(\legacy) as described in \secref{sec:back} as the compared baseline.

\begin{table}[t]
\centering
\caption{Statistics of online inference services.}
\vspace{-3ex}
\begin{tabular}{c | c c c} \hline
& Model size & \# of feature groups & Traffic load\\ \hline\hline
Service A & 430 GB & 379 & $4.58\times 10^8/s$ \\
Service B & 500 GB & 430 & $4.21\times 10^8/s$ \\
Service C & 285 GB & 270 & $3.67\times 10^7/s$ \\
Service D & 210 GB & 106 & $7.15\times 10^7/s$ \\
\hline
\end{tabular}
\label{table-data}
\vspace{-3ex}
\end{table}

In the evaluation, we focus on system performance metrics. We use latency, throughput, and the number of required instances~(\ie the computational resource) to evaluate the efficiency and cost-effectiveness of the system. Specifically, the latency is computed as the averaged time per task~(user-item pair) in millisecond~(ms), the throughput is defined as the number of tasks processed per second.
An instance refers to a virtual machine on the cloud, which is commonly equipped with four CPU cores, 20 GB memory, and shares 8 TB disk in a physical server~(4TB SSD and 4TB SATA).

\subsection{Overall Production Experience}
\tabref{table-result} reports the overall performance of \spred compared with the legacy online inference system on four different services with respect to three metrics. 
As can be seen, \spred outperforms \legacy on all services in terms of all metrics. Specifically, \spred achieves (23.33\%, 17.24\%, 2.5\%, 18.18\%) latency improvement and (188.37\%, 178.13\%, 86.16\%, 133.41\%) throughput improvement on four services, respectively. Benefit from such improvement, \spred reduces (65.33\%, 62.56\%, 46.3\%, 57.17\%) instances in the online serving infrastructure without sacrificing system efficiency, which helps Baidu save over ten million US dollars' computational resource budgets~(hardware, energy consumption, labor cost, \etc) per year. And we foresee this economic gain will improve with the future traffic load growth of each recommendation service.
Further look into the performance of each service, we notice the performance gain of Service C is relatively lower than others, which mainly because the corresponding product is resource constrained~(\ie less budget than others), and the system is tuned to minimize the resource consumption.

\begin{table}[t]
\centering
\caption{Overall performance.}
\vspace{-3ex}
\begin{tabular}{c |c| c c c c} \hline
\multicolumn{2}{c|}{}  & Latency & Throughput & \# of instance \\ \hline \hline
\multirow{2}{*}{Service A} & \legacy & 30 ms & $1.53\times 10^6$ & 11,450 \\
                         & \spred & 23 ms & $4.42\times 10^6$ & 3,970 \\
\multirow{2}{*}{Service B} & \legacy & 29 ms & $1.63\times 10^6$ & 12,750 \\
                         & \spred & 24 ms & $4.36\times 10^6$ & 4,773 \\
\multirow{2}{*}{Service C} & \legacy & 41 ms & $2.8\times 10^6$ & 2,067 \\
                         & \spred & 40 ms & $5.21\times 10^6$ & 1,110 \\
\multirow{2}{*}{Service D} & \legacy & 22 ms & $3.53\times 10^6$ & 4280 \\
                         & \spred & 18 ms & $8.24\times 10^6$ & 1833 \\\hline
\end{tabular}
\label{table-result}
\vspace{-4ex}
\end{table}

\subsection{Effect of SEDP}
Then we look into the inference time distribution to study the effect of SEDP. Due to page limit, we only report the results on Service A, the results on other services are similar.
\begin{figure}[t]
	\begin{minipage}{1.\linewidth}
    \centering
    \subfigure[Latency distribution.]{
        \includegraphics[width=.45\linewidth]{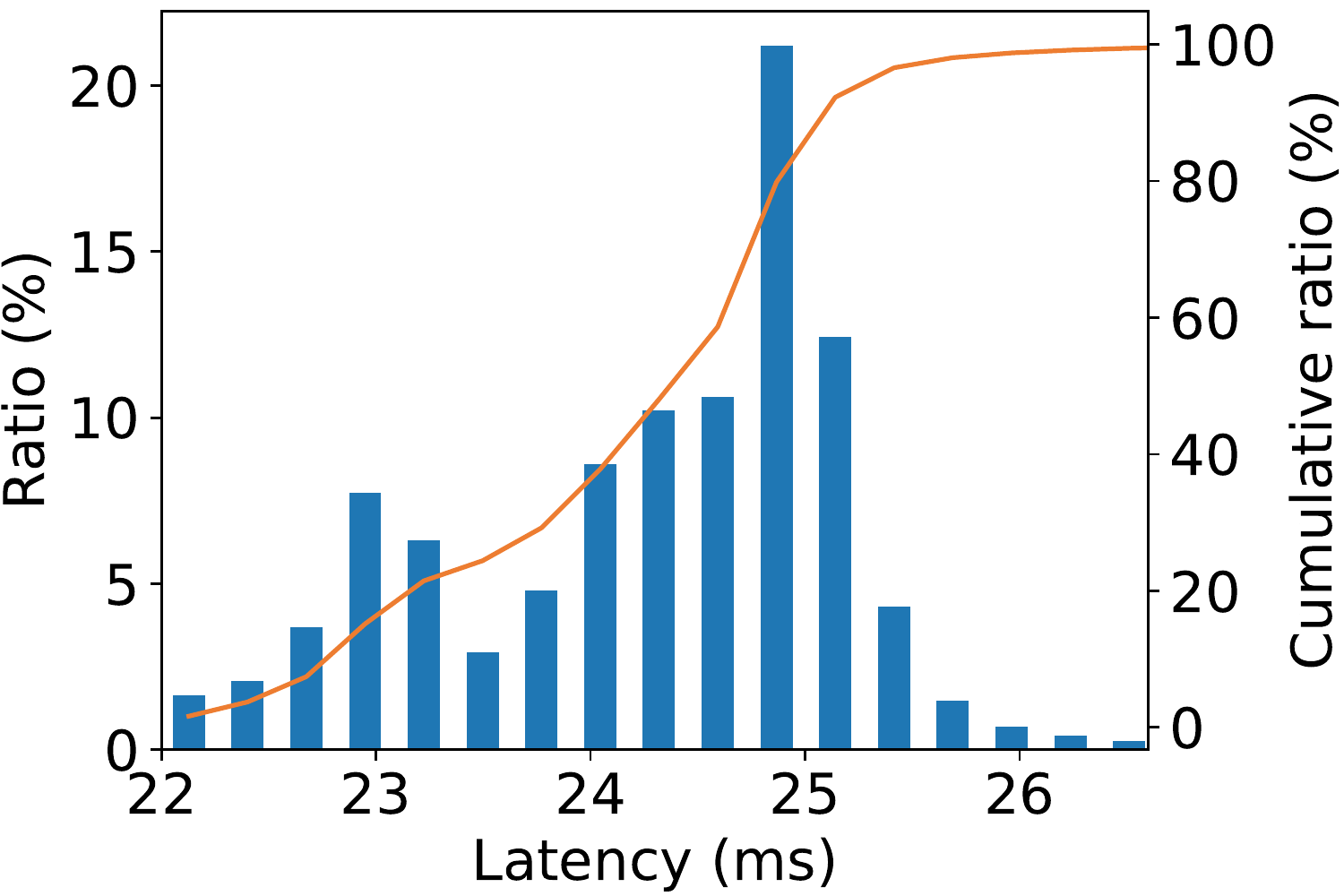}
        \label{fig:latency-dist}
    }
    \subfigure[Latency v.s. traffic.]{
        \includegraphics[width=.45\linewidth]{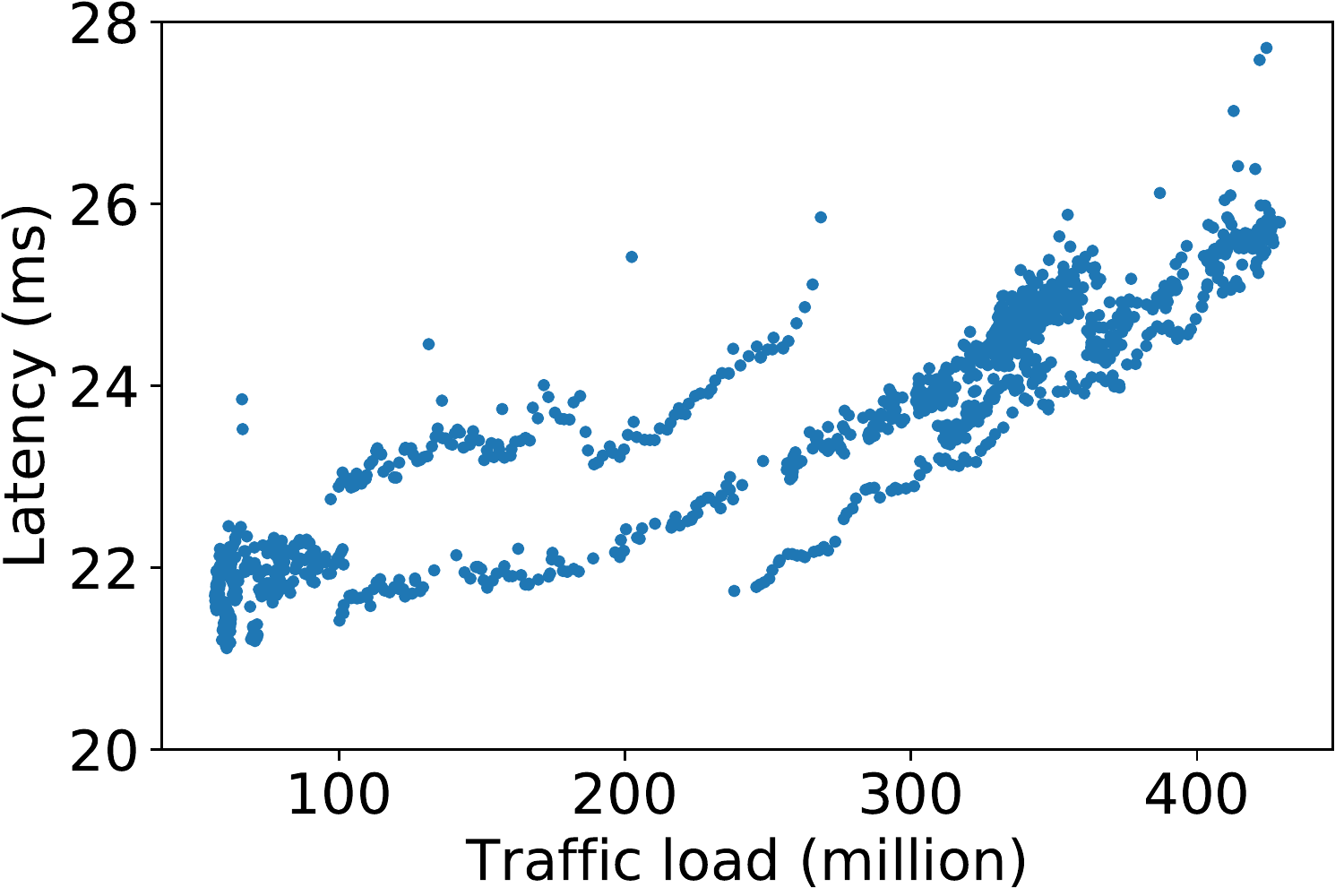}
        \label{fig:latency-traffic}
    }
    \subfigure[Latency breakdown by hour.]{
        \includegraphics[width=.45\linewidth]{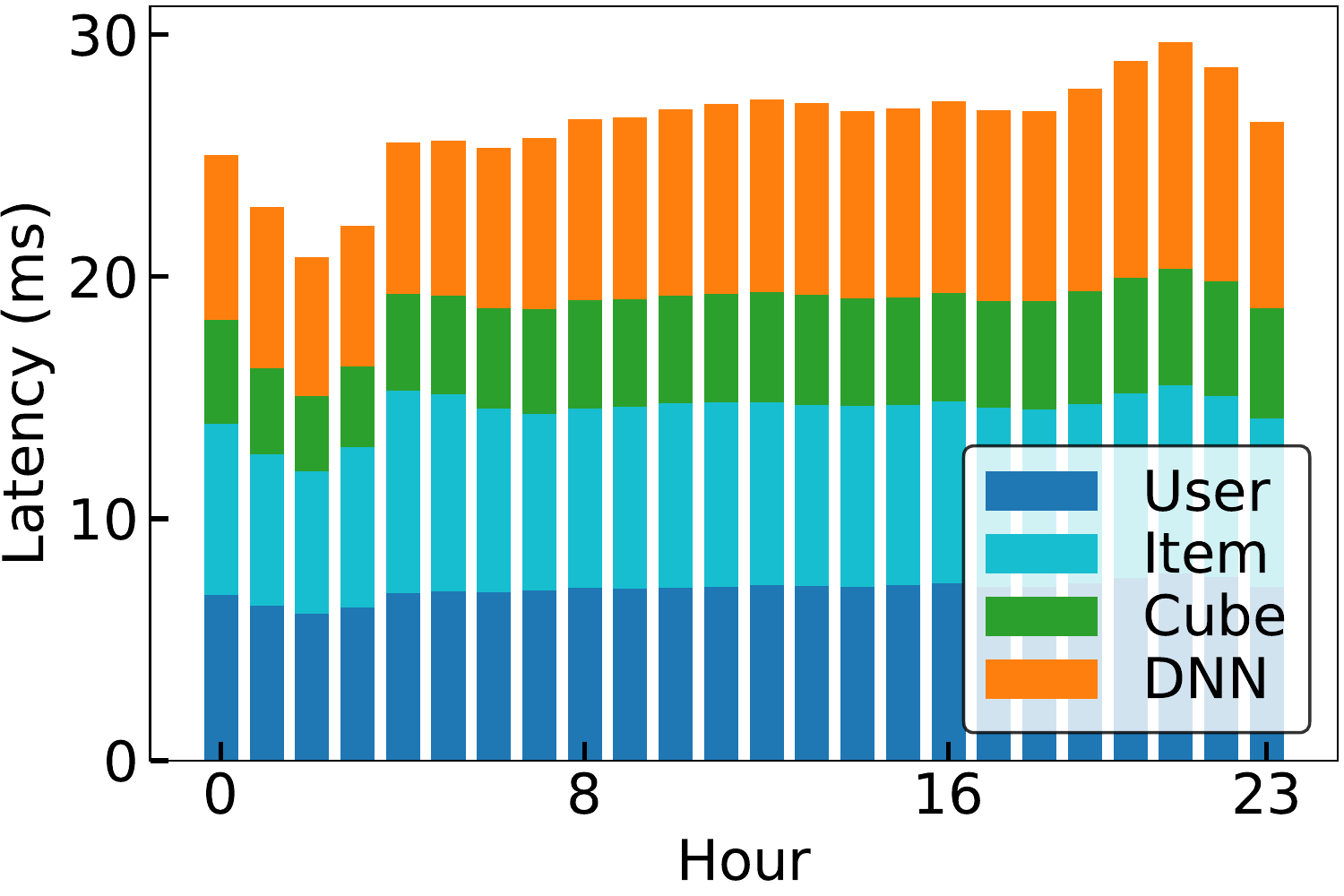}
        \label{fig:stack-latency-hour}
    }
    \subfigure[Latency breakdown by traffic.]{
        \includegraphics[width=.45\linewidth]{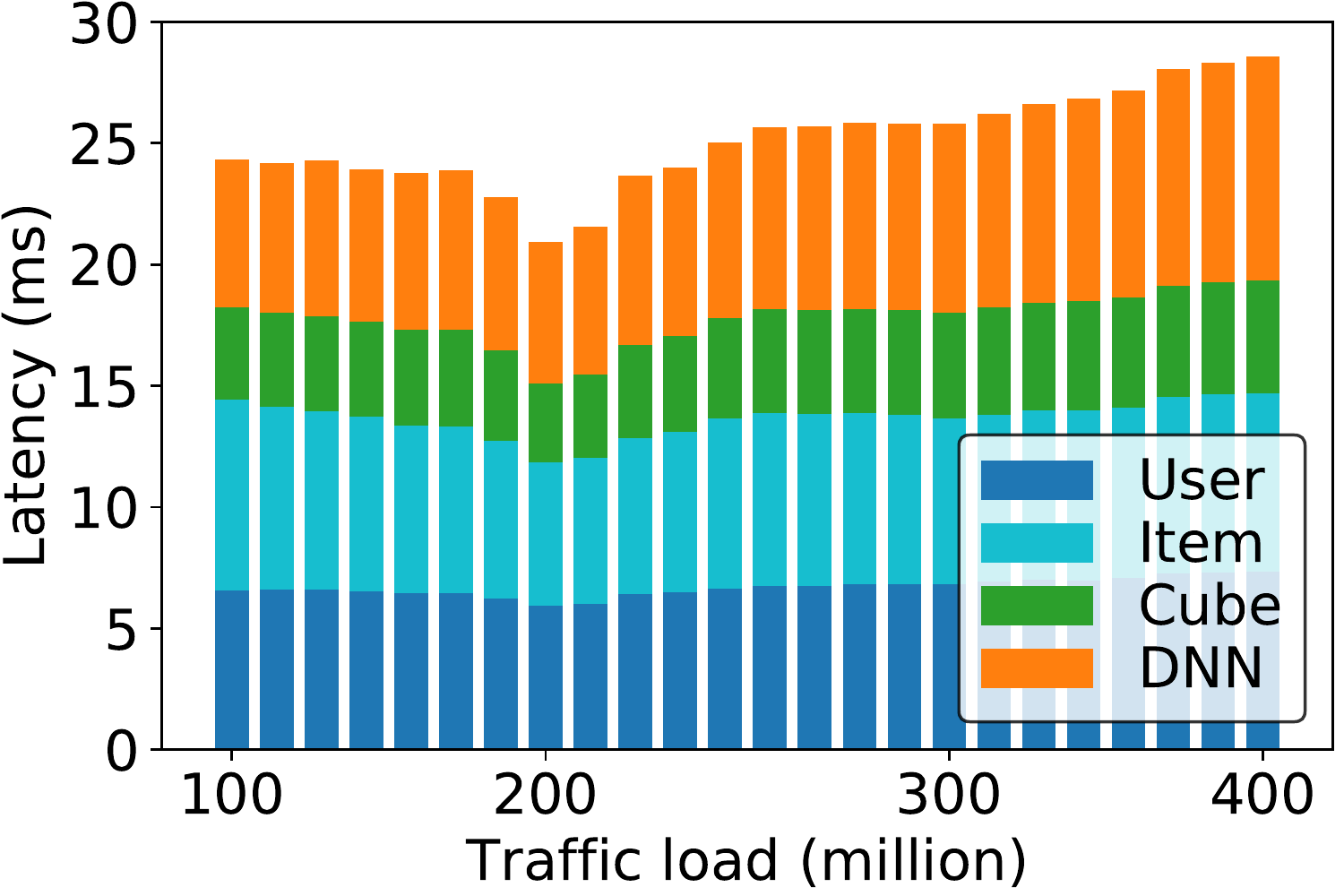}
        \label{fig:stack-latency-traffic}
    }
    \end{minipage}
    \vspace{-4ex}
    \caption{Effect of SEDP on latency.}
    \vspace{-5ex}
    \label{fig:latency}
\end{figure}

We first study the overall latency distribution.
\figref{fig:latency-dist} and \figref{fig:latency-traffic} report the overall online inference latency distribution and the latency distribution with respect to the varying traffic load, respectively.
As can be seen, the overall latency follows a bimodal distribution. This makes sense as the first peak corresponds to requests hit two caches, and the second peak corresponds to majority requests in the peak hour.
Moreover, we observe the latency increases sub-linearly with respect to the traffic load growth, which verifies the robustness of the system under time-varying traffics.

\figref{fig:stack-latency-hour} and \figref{fig:stack-latency-traffic} breakdown the latency into four major stages, \ie user related processing, item related processing, cube related processing, and DNN forward computation. 
Overall, we observe the ratio of latency in each stage is relatively stable at different hours and traffic loads.
Moreover, the overall latency slightly increases in the evening peak hour.
Besides, with the traffic load growth, we observe the averaged latency first decrease and then increase. This is related to the cache hit ratio, which first increases when the ratio of frequent requests increase and then decreases when more long-tailed items flush two caches. 

\subsection{Effect of the Storage Module}\label{sec:exp-cache}
\begin{figure}[t]
    \centering
    \includegraphics[width=.45\textwidth]{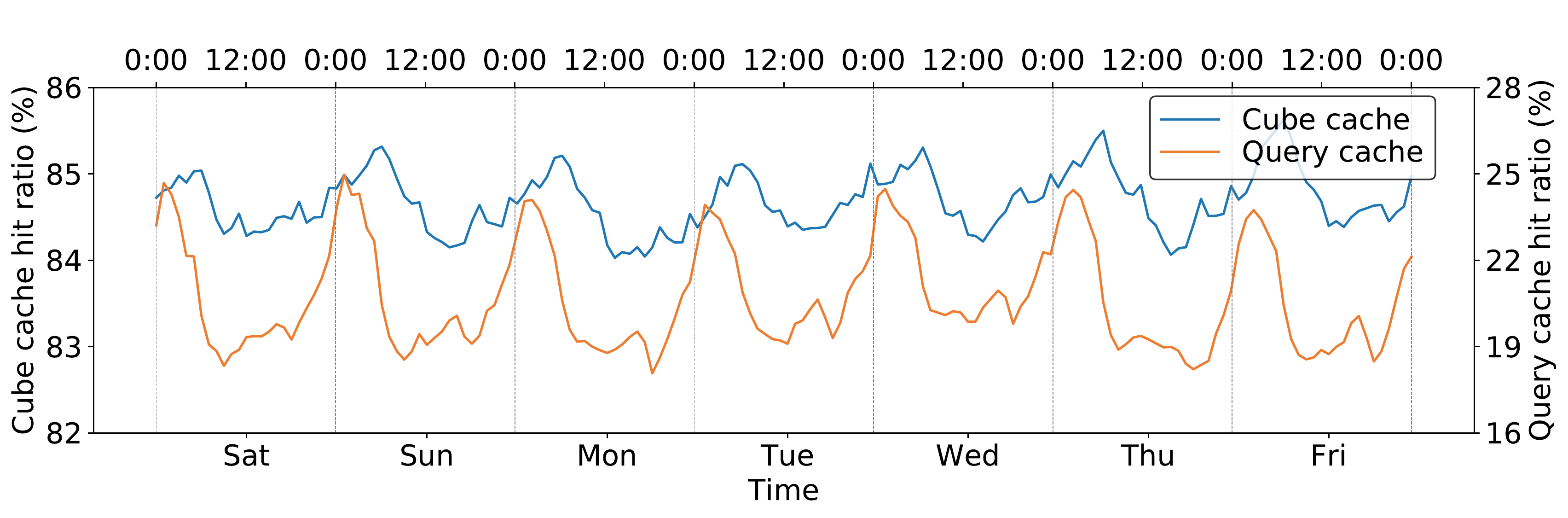}
    \vspace{-15pt}
    \caption{Hit ratio of the cube cache and query cache.}
    \vspace{-10pt}
    \label{fig:cache}
\end{figure}

We report the hit ratio of the cube cache and the query cache at different hours of Service A in a whole week to demonstrate the effect of the heterogeneous and hierarchical storage, as shown in \figref{fig:cache}. 
Overall, the hit ratio of two caches are varying by hour and showing strong daily periodicity, which is highly correlated with the traffic load variation.
In particular, the cube cache achieves $84.21\%$ hit ratio in average and the daily variation is less than $3.61\%$, while the query cache yield $19.26\%$ hit ratio in average but with a higher variation range.
In summary, the cube cache achieves higher hit ratio and every hit reduces partial data access latency, while the query cache achieves a relative lower hit ratio but per hit eliminates the whole inference task computation.
Two caches are robust to reduce the inference latency and computational overhead at different times in the week.

\begin{figure}[t]
    \centering
    \includegraphics[width=.45\textwidth]{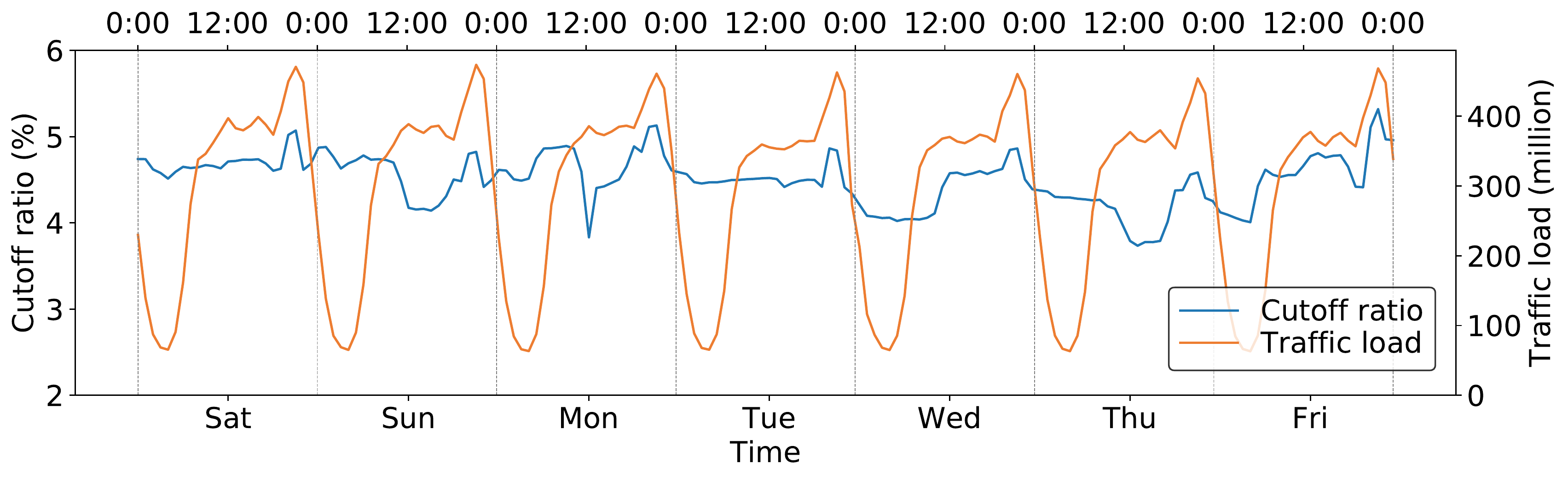}
    \vspace{-15pt}
    \caption{Cutoff ratio in Service A.}
    \vspace{-10pt}
    \label{fig:cutoff}
\end{figure}

\subsection{Effect of the Intelligent Resource Manager}
We further justify the performance gain of 
the offline resource optimization and online load shedding.

\textbf{Offline resource optimization}.
\tabref{table-offline-opt-gain} reports the quantitative results of the resource gain of each service with the offline resource optimization. 
As can be seen, the offline resource optimization module successfully reduces $8.91\%$ to $16.45\%$ instances without sacrificing system latency.
We further conduct the qualitative study by comparing key optimization parameters of Service A before~(\emph{noOpt}) and after~(\emph{Opt}) applying offline resource optimization, which is shown in \tabref{table-offline-opt-case}. As can be seen, the resource manager assigns a larger batch size to most stage processors except the cube accessor. This makes sense since over $80\%$ computations in this stage is reduced by the cube cache, as discussed in \secref{sec:exp-cache}.
Moreover, the size of the cube cache and the expire window of the query cache are increased, and the offline manager derives a more aggressive memory allocation strategy. Such settings can further reduce the data access latency and improve the CPU utilization, which therefore improves the overall resource utilization rate.

\begin{table}[t]
\centering
\caption{Resource gain of offline auto-tuning.}
\vspace{-3ex}
\begin{tabular}{c| c c} \hline
Service & Reduced \# of instance & Gain \\ \hline \hline
A & 1,636 & 14.29\% \\
B & 1,736& 13.62\%\\
C & 184 & 8.91\% \\
D & 704 & 16.45\%\\\hline
\end{tabular}
\label{table-offline-opt-gain}
\vspace{-2ex}
\end{table}

\begin{table}[t]
\centering
\caption{Offline auto-tuning results in Service A.}
\vspace{-3ex}
\begin{tabular}{c c| c | c} \hline
Category & Parameter & \emph{noOpt} & \emph{Opt} \\ \hline \hline
\multirow{5}{*}{Stage} & User batch & 30 & 34 \\
 & Item extractor batch & 4 & 12 \\
 & Item processor batch & 6 & 17 \\
 & Cube batch size & 10 & 6 \\
 & DNN batch size & 15 & 25 \\\hline
\multirow{5}{*}{System} & Cube cache ratio & 1\% & 1.2\% \\
 & Query cache window & 120s & 143s \\
 & \# of arenas & 500 & 549 \\
 & Max active extent & 6 & 25 \\
 & Huge page & Default & Always\\\hline
\end{tabular}
\label{table-offline-opt-case}
\vspace{-2ex}
\end{table}

\textbf{Online load shedding}.
To illustrate the effectiveness of online load shedding,
\figref{fig:cutoff} depicts the cutoff ratio of the online load shedding and the corresponding traffic loads over a week. 
We observe highly synchronized fluctuations of the cutoff ratio and the traffic load. Specifically, in the evening peak hour of traffic load, we observe the peak of cutoff ratio simultaneously.
Besides, we observe a relatively higher cutoff ratio at midnight, perhaps because of a lower click ratio in this period, indicating more items can be safely dropped without influencing recommendation effectiveness.

Overall, both offline and online resource management components successfully tune the online inference system to achieve better performance automatically.

\subsection{Multi-Tenant Serving}
Finally, we use \emph{Service E}, a recommendation service including multiple DNNs, to demonstrate the effectiveness of multi-tenant support in \spred. 
Specifically, \emph{Service E} including three different deep learning models, namely \emph{CTR}, \emph{FR}, \emph{CMT}, where each model corresponds to a different recommendation target.
The models are $1,743$ GB in total, involve 968 different feature groups, and the service handles $9.19\times 10^7$ traffics per second.
By deploying each different model as individual services, \tabref{table-multitenant} reports the latency, throughput and number of instances of each individual services compared with the multi-tenant based online inference on \spred.
Overall, we observe similar system latency by simultaneously executing three different models, indicating a linear speedup ratio of parallelization in \spred.
Moreover, \spred achieves $82.68\%$ throughput improvement comparing with the bottleneck model~(\ie \emph{CTR}).
Most importantly, by integrating three individual models into a unified online inference service, \spred saves $73.69\%$ instances in the cluster without losing efficiency.
\spred is even more cost-effective in the multi-tenant scenario than other single-model services~(\ie Service A to Service D).
This is because the unified service can reduce repetitive data access cost for multiple individual services~(\eg over $80\%$ feature groups are shared by three models), and largely reuses the intermediate results using the heterogeneous and hierarchical storage.

\begin{table}[t]
\centering
\caption{Effectiveness of multi-tenant serving.}
\vspace{-3ex}
\begin{tabular}{c| c c c} \hline
& Latency & Throughput & \# of instance \\ \hline \hline
\emph{CTR} & 50 ms & $2.48\times 10^6$ & 3,617 \\
\emph{FR} & 48 ms & $3.42\times 10^6$ & 2,560 \\
\emph{CMT} & 42 ms & $3.4\times 10^6$ & 1,350 \\\hline
\spred & 52 ms & $4.53\times 10^6$ & 1,980 \\\hline
\end{tabular}
\label{table-multitenant}
\vspace{-3ex}
\end{table}

\section{Conclusion and future work}\label{sec:conclusion}
In this work, we presented \spred, a fast and cost-effective online inference system at Baidu for the web-scale recommendation.
We first proposed SEDP, a directed acyclic graph style asynchronous pipeline for online inference serving.
By decoupling the inference process into modularized stage processors, SEDP reduces pipeline stall and enables flexible workflow customization and auto-tuning for a wide spectrum of recommendation scenarios.
Then we introduce the heterogeneous and hierarchical storage for cost-effective sparse DNN model management, to simultaneously accelerate the online inference speed and reduce the overall computational overhead.
After that, an intelligent resource management module is proposed to optimize the computational resource utilization in both offline and online, under inference latency constraints.
\spred has been deployed in Baidu, serving over twenty real-world internet products, and helped Baidu saved over ten million US dollars computational resource budget per year while handles hundred millions online inference requests per second without sacrificing the inference latency.
We share our design and practice experience on building the web-scale online inference system and hope to provide useful insights to communities in both academia and industry.

\bibliographystyle{ACM-Reference-Format}
\bibliography{ref}

\appendix
\section{Detailed Parameter Description}\label{appendix:extend-param}

\tabref{table-offline-param} lists some important parameters and their feasible ranges in offline tuning. Specifically, more memory related parameters can be found in the jemalloc document~(http://jemalloc.net/).
\begin{table}[t]
\centering
\caption{Representative search parameters in offline performance tuning.}
\begin{tabular}{c| c | c} \hline
 Parameter & Type & Range  \\ \hline 
\multicolumn{3}{c}{\textbf{Stage} Level Parameters}\\\hline
User batch & Integer & [10,45]  \\
 Item extractor batch &  Integer & [2,45]  \\
 Item processor batch & Integer & [2,45]\\
 Cube batch size & Integer & [1,20]  \\
 DNN batch size & Integer & [10,45] \\\hline
 \multicolumn{3}{c}{\textbf{System} Level Parameters}\\\hline
    Cube cache ratio &  continuous & [0.1,5]\% \\
    Query cache window & continuous & [60,600]s \\
    \# of arenas (jemalloc) & Integer & [350,700]\\
    Max active extent (jemalloc) & Integer & [5,40] \\
    Huge page (jemalloc) & Boolean & \{Default, Always\}\\\hline
\end{tabular}
\label{table-offline-param}
\end{table}

\tabref{table-online-feature} lists lightweight features used for intelligent online load shedding. All features are either derived from real-time system feedback or upstream processing results.
\begin{table}[t]
\centering
\caption{Features for online load shedding.}
\begin{tabular}{c| c } \hline
Feature name & Description \\ \hline \hline
quota & Available resource \\
cutoff ratio & Previous cutoff ratio \\
qid & Current item queue ID \\
escore\_avg & average of estimated scores \\
escore\_variance & variance of estimated scores \\
escore\_max & max of estimated scores \\
escore\_min & min of estimated scores \\\hline
\end{tabular}
\label{table-online-feature}
\end{table}

\begin{figure}[h]
    \centering
    \includegraphics[width=.48\textwidth]{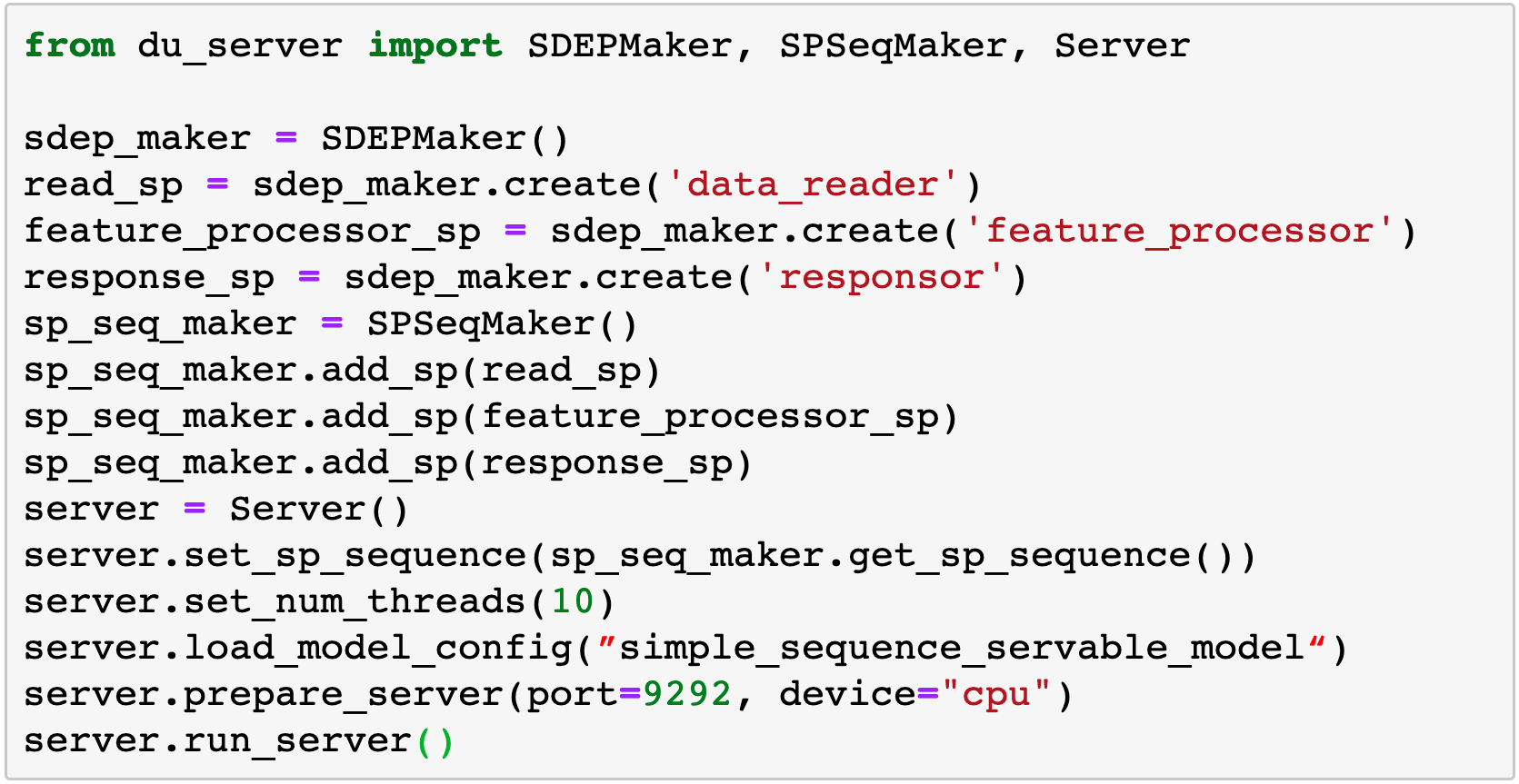}
    \caption{Constructing customized online inference service by using the Python API in a few lines of code.}
    \label{fig:code}
\end{figure}

\section{System Implementation}\label{appendix:extend-impl}

The core framework of \spred is implemented by C++.
Developers can construct customized stage processors by implementing complex data pre-possessing and post-processing logic based on several simple abstractions.
Moreover, \spred also provide a Python API extension including of a set commonly used stage processors, such as \emph{data\_reader}, \emph{feature\_processor}, \emph{cube\_accessor}, and \emph{model\_inference}.
Engineers, data scientists and researchers can easily and efficiently construct and customize online inference services for a variety of online recommendation tasks.
\figref{fig:code} showcases an example code block based on the Python API on PaddlePaddle, which defines an online inference service in a few lines of code.

\end{document}